# COMPUTER NETWORKS NEW GENERATION IN THE USE OF RES

**Kultan, J.,**
*Ekonomická univerzita v Bratislave*
jkultan@gmail.com

**Kultan, M.,**
*Slovenská technická univerzita v Bratislave*
matej.kulta@gmail.com

*Abstract*

*The paper is aimed at analyzing the potential of new information networks to solve the problems of energy management network with the use of renewable energy sources. One of the basic problems of renewable energy sources is their temporal and spatial variability. It is mainly about resources based on direct solar radiation and wind speed. New computer systems that use only classical connection-based solid structure of computer network connections but also on the basis of short-range connections allow accurate prediction of the active intensity changes observed energy. Using the system thus created can control precisely the basic energy equipment / generator and operable appliances / gradient to reduce the power needed resources or from work. This approach is one of the directions of further development of smart appliances and smart elements in the energy sector.*

## 1   INTRODUCTION

In the current period is one of the problems the use of solar energy or wind power to their volatility not only in intensity but also in space. Harmonize electricity consumption and electricity production from conventional sources and also from wind and solar power and the preparation of a reserve power at minimal cost to the energy losses to put up energy intensity changes of energy from renewable sources is a very difficult task [2].  straty energie

Despite the fact that there are many systems to weather forecast, wind speed and solar radiation, guided by their production from conventional sources in real time is a very complex process. In this process it is important not only actual parameter - energy density, but also the direction and rate of change of the parameter The figure (Fig. 7), that amendment is weather parameters on a relatively small area. For a detailed analysis, it is possible to portray the weather forecast in response to the change of parameters in the vicinity.

Analysis of these changes and their impact on the energy system is one of the preconditions for efficient use of renewable energy sources [3].

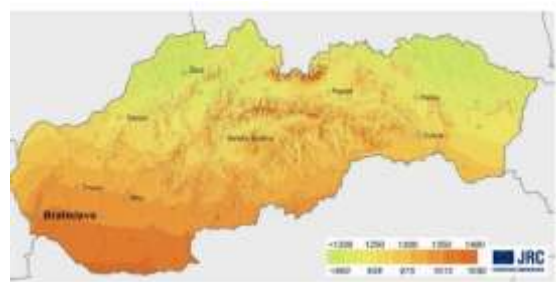

**Fig. 1 Sumárne slnečné žiarenie** http://www.kves.uniza.sk/kvesnew/dokumenty/  elektroenergetika_1_ Extern

## 2   CHANGE IN THE INTENSITY OF SOLAR RADIATION IN SPACE AND TIME

In addressing the feasibility of implementation and operation of solar - photovoltaic power plants is carried out a large number of measurements (Fig. 1) not only direct but also diffuse radiation. The disadvantage of the measurement is that taking place at the place designated for the location of the photovoltaic power station or placing solar collectors for heat recovery. In such measurements, although we have a time dependence, which serve to statistical processing possibilities of using solar energy. Such graphs, functions and calculations allow only assume with some probability on options installed power plant.

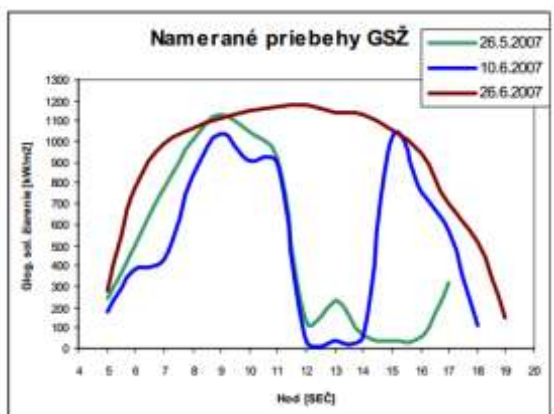

**Fig. 2 Time dependence of solar radiation during a sunny day and the days when the storm**

The intensity of solar radiation is dependent on several factors. In addition to the position of the location and angle of rotation toward the south elevation - which are heavily influenced by variables do not change over time and other factors that are time-dependent, ie slope depending on the particular period of the day and year, and random - cloud, and speed of movement of clouds dependent on the wind speed and direction.

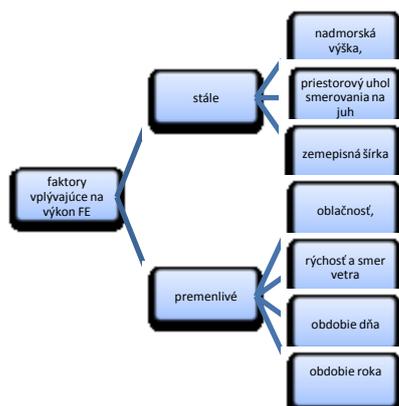

**Fig. 3 Factors affecting the change in performance FE**

For controlling the operation of the energy system is necessary to know not only the degree of cloud in the region of the installed power plant (Fig. 7), but also the speed of the cloud systems. In the case of full power plant is a risk that at some point the wind drifts clouds that reduce power output power, or vice versa, due to fluctuations in the power plant cloud gradually increasing.

Total energy in the network due to changes in cloud affects not only the power plant itself, but also news consumers in the near and far the cloud.

Assume that storm clouds converge at the place of the curve (Fig. 4) $\boldsymbol{k}$(A (x, y), r). At this point, reduce the lighting and begins to increase of energy consumption in a given area.

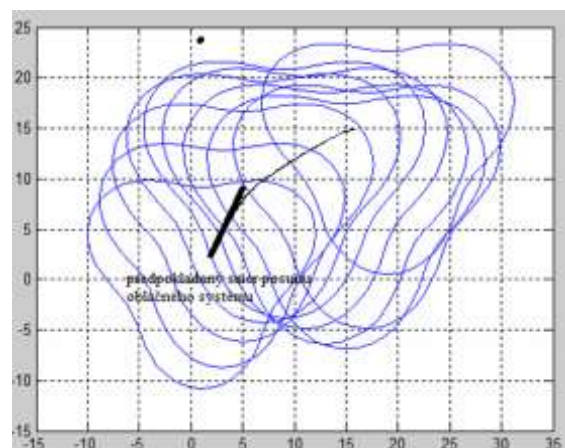

**Fig. 4  Redistribute the rain area**

Due to the reduction of solar radiation begins to decrease in temperature which may result in an increase of energy consumption or for heating / winter season / or a sharp decrease due to a reduction in sampling subscriptions conditioning / in the summer period /. It also runs photovoltaic power power reduction due to the reduction of direct sunlight. These changes are not static, but gradually taking place in the direction of the main flow direction (Fig. 5) and result in a gradual increase in the concentration of cloud - where the direction of movement are emerging hazards, or reduce the concentration of cloud - for enlargement cloudy ring flat in free space.

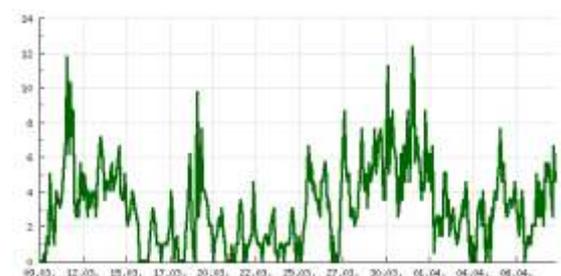

**Fig. 5 Time dependence of the intensity of the wind sweeping down storm clouds**

In classical measurement of the intensity distribution of radiation intensity we get a function of time, the above method we obtain more precise data on the rate of change of intensity of radiation from other variables / wind speed, direction, eventually the pressure ... /.

$$\Delta E = \boldsymbol{k}S \int_{t_0}^{t_1} I(t, v, p, ..) \, dt$$

The equation represents the change in the calculation of energy supplied in response to the change in radiation intensity over time, where k is the coefficient transformation process involving radiation intensity on the acquired energy

Even in this case, the sensors tend stored at the site localization of power, but the accuracy of forecasting intensity changes is much higher.

The advantage of the proposed solution is the possibility of creating maps of changes in the overall energy balance depending on the weather (intensity of radiation, cloud, speed and direction of wind flow, pressure, temperature, ...), field, periods of day or season.

## 3  THE PREDICTION OF CHANGE IN THE INTENSITY OF SOLAR RADIATION USING COMMUNICATION SYSTEMS

When using RES is one of their weaknesses variability in the intensity / energy density / and time. Such uncertainty makes it difficult to fully deployed renewable energy production. Taking into account the constant change of energy consumption, to ensure the quality and stable supply is necessary to use a large amount of additional energy, eg. in the form of hot reserves in conventional power plants [4]. Losses arising from uncertainty sampling and production can be reduced, provided not only improving predictions of energy production from renewable sources, but also forecasts a change in consumption due to changes in weather conditions at any given time.

Often used statistical models allow to estimate the necessary energy balance with some probability, but not a direct answer to the question: "How to change control of generators conventional power plants to adapt to changing weather conditions."

### 3.1 Smart Sensors weather conditions

In managing energy system using renewable energy sources, we propose to install a large number of smart devices that measure not only the state of the network but also state temperature - temperature, pressure, humidity, intensity of solar radiation, lighting and other parameters. Consequently, these variables are evaluated and can send data outside the appliance. Based on the comparison of the values of wind speed and direction data that came from other smart devices, it is possible to determine the direction and velocity. For measurements of solar radiation and comparison with those of other smart devices, we can compare the movement of individual cloud systems. The basic prerequisite for the creation of such systems is to establish channels of communication between the smart elements rather than through traditional communication networks based on hierarchical level of interconnected switches and routers, but through channels direct connection of individual elements (Fig. 6). Such systems use signals to communicate short range - wife, Bluetooth and other.

Communication then takes place very quickly and on the basis of location information of each smart device and intensity of the measured quantity is possible to create an accurate model of the dynamic changes not only the movement but also the intensity of the measured variable - sunshine.

In addition to the condition of the neighborhood, the smart device senses the state of the network [1]. The overall importance of these devices lies in creating a data set designed for creating dynamic model of network status and condition of the environment under which it is possible to relatively accurately design algorithm changes the fundamental sources placed in thermal power plants. Considering that smart devices are located in different parts of the electricity network and mutual communication is very fast due to the arrangement, it obtained the overall dynamic model provides enough data to obtain information on changes in the network.

### 3.2 Local management through smart energy network elements

In this case, it is not a total global energy management system at the level of control rooms, or management of individual plants. The proposed governance model is to reduce unforeseen dynamics of changes in the performance of individual sources close to failure prompted the change of solar power plants. Assuming that this is a change in performance by 10-40% in some cases - strong storm activity with high wind speeds - and even more installed power plant is 1MW, such dropout / grad / has an effect on the next power mode. There is an unforeseen change in power plants. In the case of short-term changes have only power losses due to the increase in the flow of production, in the event that it is a long-term changes - there is economic losses due to the need to purchase additional energy to meet these changes.

As a result, it is necessary that smart equipment / changing their collection or production, depending on the network conditions / quickly responded to the said condition and able to manage its activities with some advance.

In previous posts was decommissioned examples of the smart network elements depending on the current condition of the network. In this case, it is assumed that smart elements will react to the weather conditions, forecasts and development capabilities to respond to possible changes in power sources. In the introductory part mentioned that weather change also affects the change in consumption curve, especially in the transition period or during the summer storms. Said devices may be necessary helper especially in summer storm season, when storm clouds move relatively high speed, obscure individual solar power and turn them odokrývajú while increasing ordering by increasing lighting in buildings. When leaving clouds of this phenomenon occurs in the opposite direction.

Take advantage of smart elements of the energy system will react to the changes much faster and better than management systems for individual plants. It should also be noted that the propose load curve, taking into account such changes receive the network is virtually impossible.

### 3.3 Communication Elements and smart energy management system

A large part of the solution of energy problems of optimal control system using a large number of smart features lies in addressing mutual communication between individual elements .

The use of traditional communication networks based on computer networks and hierarchical connections through routers, switches , high-speed internet on metallic or optical fibers is probably not the appropriate solution to the problem. Although the speed of data transfer from one place to another in such a system is relatively large, in this case it is a small package transfer data between a large number of participants . While even those participants may not be geographically far away , their mutual connection can be organized through a complex system of network connections . Moreover, the number of elements that communicate with each other is very large and required data rate and data between elements of high growth , depending on the approach of the individual elements together. If we have N devices that communicate with each other and exchange , for example . Mi , j the amount of data that the total amount of data transmitted in the central node is the sum of the amount of data transferred from point i to point j.

$$M = \sum_{i=1}^{n} \sum_{j=1}^{n} M_{i,j}$$

That relationship is especially true for networks with one central node. This node must handle all calls, and the total amount of data derive the technical requirements for a given node.

For our network we can replace that relationship communication devices associated with one access point.

Relation holds for the router

$$M = \sum_{i=1}^{n_1} \sum_{j=1}^{n_2} M_{i,j}$$

where $n_1$ and $n_2$ is the number of elements in the first and the second subnet.

Also, if the element i is from the set of the first subnet, j is an element of the plurality of second subnet.

On this basis, the activities of smart network elements is essential other means of communication between them. The most obvious is the use of short range devices and mutual transfer of data between adjacent elements. Another situation occurs in networks with a direct link. In this case, we assume that each device can be connected to another device with a wireless connection / wifi, blutooth, infrared connection , etc../ For the purposes of existing networks , we propose using wifi connection especially having sufficient impact . Thus ended the network does not have a mutual connection through a common access point , but connects each element to each precisely when the connection is necessary . The speed of such a process of progressive transmission of data on changes in intensity to each element of the network is similar to the rate of change of network status from the epicenter of such changes . Innovation in the way we communicate is the case when several LANs close together so that some one network devices are within range of the other . In this case it is not necessary that the networks were linked by a router that is wifi - wifi router - but close to one network elements transmit information to the other network elements other ( Fig. 6 ). Multichannel transmission occurs , the rate of which is determined by linking each device . From communication may be excluded transmission **element - wifi - router - wifi - element** and replace only the transmission **element - elemen**t .

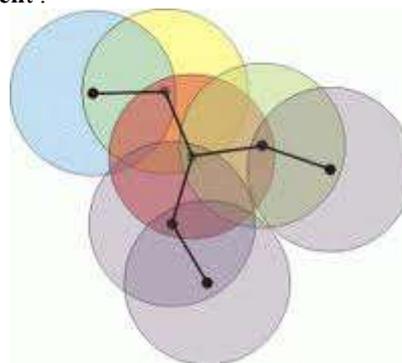

**Fig. 6 Transition signal from the smart elements excluding transition hierarchical networks**

Considering that smart elements while responding to the information obtained on the spread of the network changes, and actually there is a rapid loss of this change and its spread is significantly smaller than without these smart features.

## 4    CONCLUSION

The current energy networks with the use of renewable energy sources is a system that is highly dependent on the nature of renewables. To reduce the impact and maintain the stability and quality of energy supplied shall be interpreted a lot of energy, which reduces energy and economic efficiency of the utilization of the resources.

With a view to reducing the discrepancy between free and expensive energy source is necessary to introduce new advanced technologies to his speed response not only to the state but also the entire network near and far and the speed of communication allow dynamically manage the entire network with the use of minimal changes classical sources not only but in terms of time and performance. The network itself as a large system has sufficient internal reserves to allow short change the power or energy flow so as to eliminate short-term changes. Currently being addressed by the elimination of the increase or decrease in the level of performance of thermal and hydroelectric power plants or buying and selling energy, which is economically challenging and has resulted in an increase in overall energy prices.

One way is to develop new features smart energy networks and develop a new generation communication networks.

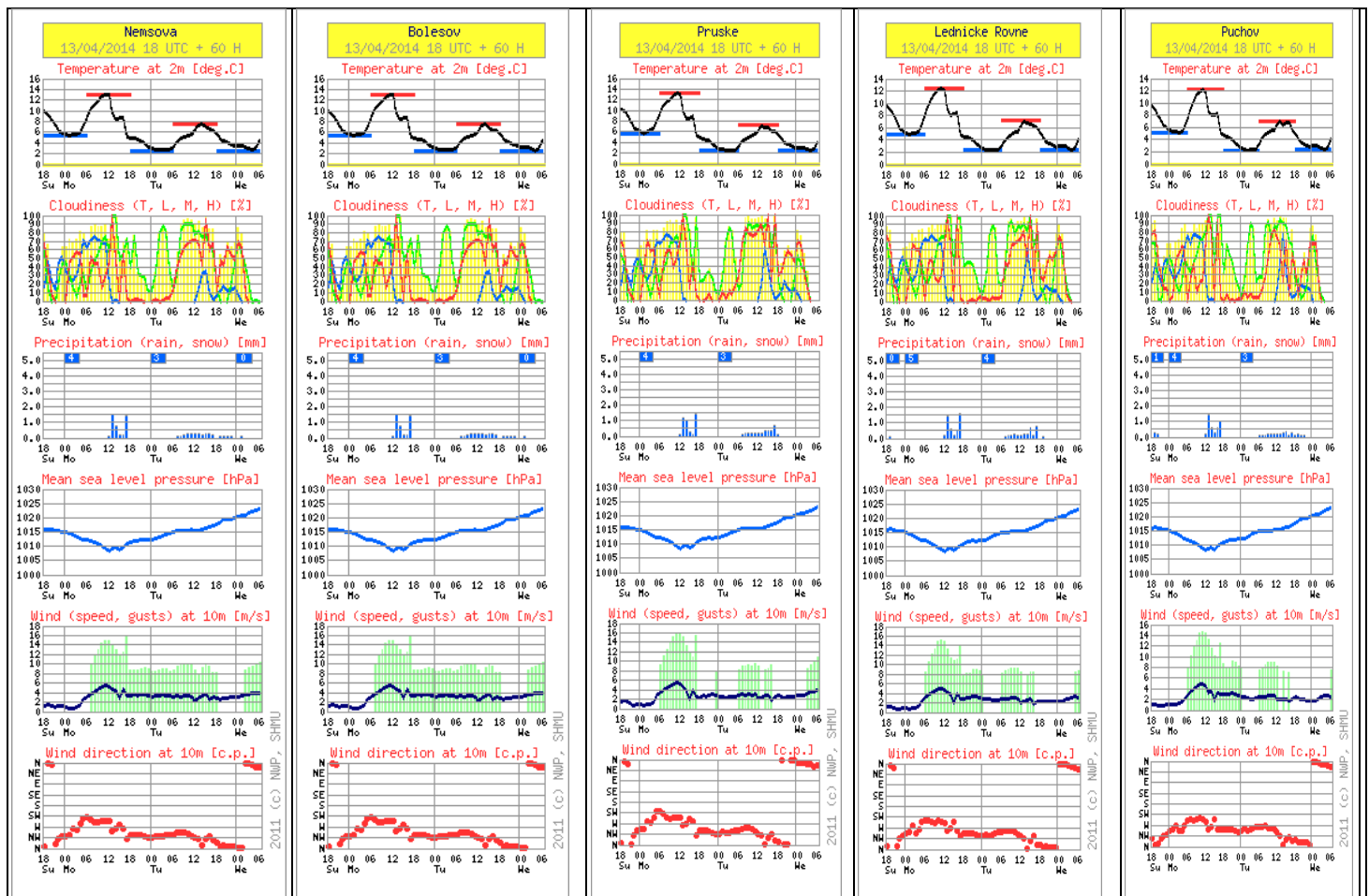

**Fig. 7   Changing weather parameters in the nearby areas of electricity consumption**

## 5   ACKNOWLEDGEMENTS

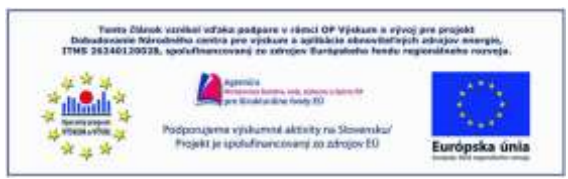

## 6   REFERENCES

,

[1] ] KULTAN, J.: Obnoviteľné zdroje energie a riadenie spotrebičov - využitie smart technológií v energetike = Renewable energy resources and management appliances - use of smart Technologies in the energy In Energy - Ecology - Economy 2012 : proceedings of the 11th international scientific conference : Tatranské Matliare, Slovakia, May 15-17, 2012. - Bratislava : Slovak University of Technology in Bratislava, 2012. - ISBN 978-80-89402-50-2. - S.[1-9].

[2] KULTAN, J.:Vybrané aspekty využívania obnoviteľných zdrojov energie (Selected Aspects Of Renewable Energy Sources Exploitation), 8 celoštátna konferencia s medzinárodnou účasťou Energetika Ekológia Ekonomika 2009, Vysoké tatry, 27.-29. mája 2009, ISBN 978-80-89402-08-3

[3] KULTAN, J.: Model trhu s elektrinou. Ekonomické aspekty výroby, prenosu a distribúcie elektriny v Slovenskej Republike, STU v Bratislave, 2009, Číslo ISBN 978-80-89402-10-6; BAB

[4] JANÍČEK, F. a kol.: Dopady vplyvu nárastu výroby elektriny z Obnoviteľných zdrojov

energie (OZE) vyvedených do distribučných sústav na prevádzkovateľa PS a účastníkov trhu s elektrinou, Slovenská technická univerzita v Bratislave, Fakulta elektrotechniky a informatiky, Katedra elektroenergetiky, Zmluva o dielo: 41/130/2008

[5] JANÍČEK, F., KULTAN, J., KOREC, M., ŠEDIVÝ, J., ŠULC, J.: Obnoviteľné zdroje energie v podmienkach SR, Elektrotechnika, Informatika a telekomunikácie, Časopis pre elektrotechniku a energetiku ročník 14, október 2008, mimoriadne číslo, str. 148-155